\documentstyle[12pt,aasms4]{article}

\makeatother
\def\eq#1{\begin{equation} #1 \end{equation}}

\def\case#1/#2{\hbox{$\frac{#1}{#2}$}}

\def\comment#1        {\tt #1}
\def\refto#1          {\ref #1}

\begin{document}

\rightline{\it Accepted by Astronomical Journal}

\title{             Color Confirmation of Asteroid Families    }

\author{
\v{Z}eljko Ivezi\'{c}\altaffilmark{\ref{Princeton}},
Robert H. Lupton\altaffilmark{\ref{Princeton}},
Mario Juri\'{c}\altaffilmark{\ref{Princeton},\ref{Zagreb},\ref{Visnjan}},
Serge Tabachnik\altaffilmark{\ref{Princeton}},
Tom Quinn\altaffilmark{\ref{Washington}},
James E. Gunn\altaffilmark{\ref{Princeton}},
Gillian R. Knapp\altaffilmark{\ref{Princeton}},
Constance M. Rockosi\altaffilmark{\ref{Washington}},
Jonathan Brinkmann\altaffilmark{\ref{APO}}
}

\newcounter{address}
\setcounter{address}{1}
\altaffiltext{\theaddress}{Princeton University Observatory, Princeton, NJ 08544
\label{Princeton}}
\addtocounter{address}{1}
\altaffiltext{\theaddress}{University of Zagreb, Dept. of Physics, Bijeni\v{c}ka 
cesta 32, 10000 Zagreb, Croatia
\label{Zagreb}}
\addtocounter{address}{1}
\altaffiltext{\theaddress}{Vi\v{s}njan Observatory, Istarska 5, HR-52463 Vi\v{s}njan,
Croatia
\label{Visnjan}}
\addtocounter{address}{1}
\altaffiltext{\theaddress}{University of Washington, Dept. of Astronomy,
Box 351580, Seattle, WA 98195
\label{Washington}}
\addtocounter{address}{1}
\altaffiltext{\theaddress}{Apache Point Observatory,
2001 Apache Point Road, P.O. Box 59, Sunspot, NM 88349-0059
\label{APO}}

\begin{abstract}
We discuss optical colors of 10,592 asteroids with known orbits selected from a
sample of 58,000 moving objects observed by the Sloan Digital Sky Survey (SDSS). 
This is more than ten times larger sample that includes both orbital parameters and 
multi-band photometric measurements than previously available. We confirm that asteroid 
dynamical families, defined as clusters in orbital parameter space, also strongly 
segregate in color space. In particular, we demonstrate that the three major asteroid 
families (Eos, Koronis, and Themis), together with the Vesta family, represent four 
main asteroid color types. Their distinctive optical colors indicate that the
variations in chemical composition within a family are much smaller than the 
compositional differences between families, and strongly support earlier suggestions 
that asteroids belonging to a particular family have a common origin. We estimate
that over 90\% of asteroids belong to families.
\end{abstract}

\section{                       Introduction               }

Asteroid dynamical families are groups of asteroids in orbital element
space (Gradie, Chapman \& Williams 1979, Gradie, Chapman \& Tedesco 1989, 
Valsecchi {\em et al.} 1989). This clustering was first discovered by Hirayama 
(1918, for a review see Binzel 1993), who also proposed that families 
may be the remnants of parent bodies that broke into fragments. About half 
of all known asteroids are believed to belong to families; recent work 
(Zappal\'{a} {\em et al.} 1995, hereafter Z95), applying a hierarchical clustering method
to a sample of 12,487 asteroids, finds over 30 families. The contrast between
families and the background is especially strong in the space spanned by the so-called
{\it proper} orbital elements. These elements are nearly invariants of motion
and are thus well suited for discovering objects with common dynamical history 
(Valsecchi {\em et al.} 1989, Milani \& Kne\v{z}evi\'{c} 1992, hereafter MK92).
The current asteroid motion is described by {\it osculating} orbital
elements which vary with time due to perturbations caused by planets, and are
less suitable for studying dynamical families.

Asteroid clustering is much weaker in the space spanned by directly observed
osculating elements than in the space spanned by derived proper elements.
Figure 1 compares the osculating (top panel, Bowell 2001) and proper
(bottom panel, MK92) orbital inclination vs. orbital
eccentricity distributions of 1,720 asteroids from the outer region of the main
asteroid belt (proper semi-major axis larger than 2.84 AU).
This region contains all three major asteroid families:
Eos, Koronis and Themis, with approximate ($a, \sin(i), e$) of (3.0, 0.18, 0.08), (2.9,
0.03, 0.05) and (3.15,0.02, 0.15), respectively. Here $a$ is proper semi-major axis,
$\sin(i)$ is the sine of the orbital inclination angle, and $e$ is eccentricity.

The proper elements are derived from the osculating elements by an approximate 
perturbation method (MK92), and it is possible that the overdensities evident 
in the bottom panel are at least partially created by that algorithm (Valsecchi 
{\em et al.} 1989, Bendjoya 1993). A firm proof that families are real therefore requires 
their confirmation by a method that is {\it not} based on dynamical considerations, 
for example, that dynamically selected groups have distinctive
colors. While there is observational evidence that at least the most populous asteroid
families have characteristic colors (Degewij, Gradie \& Zellner 1978, Chapman 1989), 
even the most recent studies of the colors of asteroid families include fewer than 50 
objects per family (Florczak {\it et al.} 1998, Doressoundiram {\it et al.} 1998, 
Florczak {\it et al.} 1999). The large number (about 10,000) of color measurements 
for catalogued asteroids (Bowell 2001) recently made available by the Sloan Digital Sky
Survey (SDSS, York {\it et al.} 2000) allows a detailed investigation of this question.

\section{                      SDSS Observations of Asteroids              }

The SDSS is a digital photometric and spectroscopic survey which will cover one quarter
of the Celestial Sphere in the North Galactic cap and produce a smaller but much deeper
multi-epoch survey in the Southern Galactic hemisphere (Stoughton {\it et al.} 2002). 
The survey sky coverage will result in photometric measurements (Smith {\it et al.} 2002, 
Hogg {\it et al.} 2002) for about 50 million stars and a similar number of galaxies,
and spectra for about 1 million galaxies and 100,000 quasars. Although primarily designed
for observations of extragalactic objects, the SDSS is significantly contributing to studies
of solar system objects, because asteroids in the imaging survey must be explicitly
recognized to avoid contamination of the quasar samples selected for spectroscopic
observations (Lupton {\it et al.} 2001). The SDSS will increase the number of asteroids with 
accurate five-color photometry (Fukugita {\it et al.} 1996, Gunn {\it et al.} 1998)
by more than two orders of magnitude (to about 100,000), and to a limit more than five 
magnitudes fainter than previous multi-color surveys (Ivezi\'{c} {\em et al.} 2001,
hereafter I01).

\subsection{                       SDSS Moving Object Catalog      }

Most of the asteroids observed by the SDSS are new detections, because the SDSS finds
moving objects to a fainter limit ($V\sim21.5$) than the completeness limit of currently
available asteroid catalogs ($V\sim18$). However, SDSS observations, which are obtained
with a baseline of only 5 minutes (Lupton {\em et al.} 2001, I01), are insufficient to 
determine accurate orbits, and we consider only objects that have previously determined 
orbital parameters. The details of the matching procedure and a preliminary sample are 
described by Juri\'{c} {\em et al.} 2002 (hereafter J02). Here we extend their analysis 
to a significantly larger sample, and introduce a new method for visualizing the 
distribution of asteroids in a multi-dimensional space spanned by orbital parameters 
and colors.  

The currently available SDSS moving object list (Ivezi\'{c} {\em et al.} 2002, hereafter
SDSSMOC) includes over 58,000 observations; 10,592 are detections of unique objects listed 
in the catalog of known asteroids (Bowell 2001), and 2,010 detections are multiple 
observations of the same objects. For a subset of 6,612 objects from this list, the 
proper orbital elements are also available (MK92) and are analyzed here.  These samples are 
about an order of magnitude larger than used in previous studies of the colors of asteroids, 
and also benefit from the wide wavelength range spanned by SDSS filters\footnote{The $z$
band extends to the near-infrared range and allows efficient recognition of Vesta type
asteroids (Binzel \& Xu 1995).} (Gunn {\it et al.} 1998).

\subsection{                Asteroid Colors as Observed by SDSS    } 

SDSS colors can distinguish asteroids of at least three different color types
(I01, J02). Using four of the five SDSS bands, we construct the color-color diagram shown 
in Figure 2. The horizontal axis\footnote{See I01 for a discussion of $a^*$ color.} is
\eq{
\label{acolor}
            a^* \equiv 0.89 \,(g - r) + 0.45 \,(r - i) - 0.57,
}
and the vertical axis is $i-z$, where $g-r$, $r-i$, and $i-z$ are the 
asteroid colors measured by SDSS (accurate to about 0.03 mag). Each dot represents
one asteroid, and is color-coded according to its position in this diagram (note that these 
colors do not correspond directly to asteroid colors as would be seen by the human eye). 
As discussed by I01, the asteroid distribution in this diagram\footnote{For the position
of asteroid taxonomic classes in this diagram see Figure 10 in I01.} is highly bimodal, with 
over 90\% of objects found in one of the two clumps that are dominated by rocky S type 
asteroids ($a^*\sim0.15$), and carbonaceous C type asteroids ($a^*\sim-0.1$). Most of 
the remaining objects have $a^*$ color similar to S type asteroids, and distinctively 
blue $i-z$ colors. They are dominated by Vesta type asteroids (Binzel \& Xu 1995, 
J02).

Figures 3 and 4 show two two-dimensional projections of the asteroid distribution in the 
space spanned by proper semi-major axis, eccentricity, and the sine of the orbital inclination 
angle, with the points color-coded as in Figure 2. The vertical bands where practically no 
asteroids are found (at $a$ of 2.065, 2.501, 2.825 and 3.278 AU) are the 4:1, 3:1, 5:2, and 
2:1 mean motion resonances with Jupiter (the latter three are the Kirkwood gaps). Figure~5 
is analogous to the bottom panel in Figure~1.

\section{                              Discussion                  }

A striking feature of Figures 3, 4 and 5 is the color homogeneity and distinctiveness displayed
by asteroid families. Each of the three major Hirayama families, Eos, Koronis and Themis, and
also the Vesta family at ($a, \sin(i), e$) of (2.35, 0.12, 0.09), has a
characteristic color. This strong color segregation provides firm support for the reality of
asteroid dynamical families. The correlation between the asteroid colors and their heliocentric
distance has been recognized since the earliest development of asteroid taxonomies
(Chapman, Morrison \& Zellner 1975, Gradie \& Tedesco 1982, Zellner, Tholen \& Tedesco 1985,
Gradie, Chapman \& Tedesco 1989). Our analysis indicates that this mean correlation (see
e.g. Figure 23 in I01) is mostly a reflection of the distinctive colors of asteroid families 
and their heliocentric distribution.

When only orbital elements are considered, families often partially overlap each
other (Z95), and additional independent information is needed to 
improve their definitions. With such a massive, accurate and public database as that discussed 
here (SDSSMOC), it will be possible to improve the classification of asteroid families by
simultaneously using both the orbital elements and colors. For example, the SDSS colors
show that the asteroids with ($a, \sin(i)$) about (2.65, 0.20) are distinctively blue (Figure 3),
proving that they do not belong to the family with ($a, \sin(i)$) about (2.60, 0.23), but
instead are a family in their own right. While this and several similar examples were already
recognized as clusters in the orbital parameter space (Z95), this work
provides a dramatic independent confirmation. Figures 3, 4 and 5 suggest that the asteroid
population is dominated by families: even objects that do not belong to the most populous
families, and thus are interpreted as background in dynamical studies, seem to show color
clustering. Using the definitions of families based on dynamical analysis (Z95),
and aided by SDSS colors, we estimate that at least 90\% of asteroids are associated with
families\footnote{The preliminary analysis indicates that about 1--5\% of objects do
not belong to families. A more detailed discussion of the robustness of this result will 
be presented in a forthcoming publication. Similarly, it is not certain yet whether 
objects not associated with the families show any heliocentric color gradient.}.

Proper orbital elements (MK92) are not available for asteroids with large
semi-major axis and orbital inclination. In order to examine the color distribution
for objects with large semi-major axis, such as Trojan asteroids ($a\sim 5.2$) and
for objects with large inclination, such as asteroids from the Hungaria family ($a\sim 1.9,
\sin(i)\sim 0.38$), we use osculating orbital elements. Figure 6 shows the distribution
of all the 10,592 known asteroids observed by the SDSS in the space spanned by osculating
semi-major axis and the sine of the orbital inclination angle, with the points
color-coded as in Figure 2. It is remarkable that various families can still be easily
recognized due to SDSS color information. This figure vividly demonstrates that the asteroid
population is dominated by objects that belong to numerous asteroid families.

\vskip 0.4in
\leftline{Acknowledgments}

We are grateful to E. Bowell for making his ASTORB file publicly available, and to
A. Milani, Z. Kne\v{z}evi\'{c} and their collaborators for generating and distributing
proper orbital elements. We thank Princeton University for generous financial support 
of this research, and M. Strauss and D. Schneider for helpful comments.

The Sloan Digital Sky Survey (SDSS) is a joint project of The University of Chicago,
Fermilab, the Institute for Advanced Study, the Japan Participation Group, 
The Johns Hopkins University, the Max-Planck-Institute for Astronomy (MPIA), 
the Max-Planck-Institute for Astrophysics (MPA),
New Mexico State University, Princeton University, the United States Naval Observatory, and the
University of Washington. Apache Point Observatory, site of the SDSS telescopes, is operated by
the Astrophysical Research Consortium (ARC). Funding for the project has been provided by the
Alfred P. Sloan Foundation, the SDSS member institutions, the National Aeronautics and Space
Administration, the National Science Foundation, the U.S. Department of Energy, the Japanese
Monbukagakusho, and the Max Planck Society. The SDSS Web site is http://www.sdss.org/.


\newpage

\clearpage

\begin{figure}
\plotfiddle{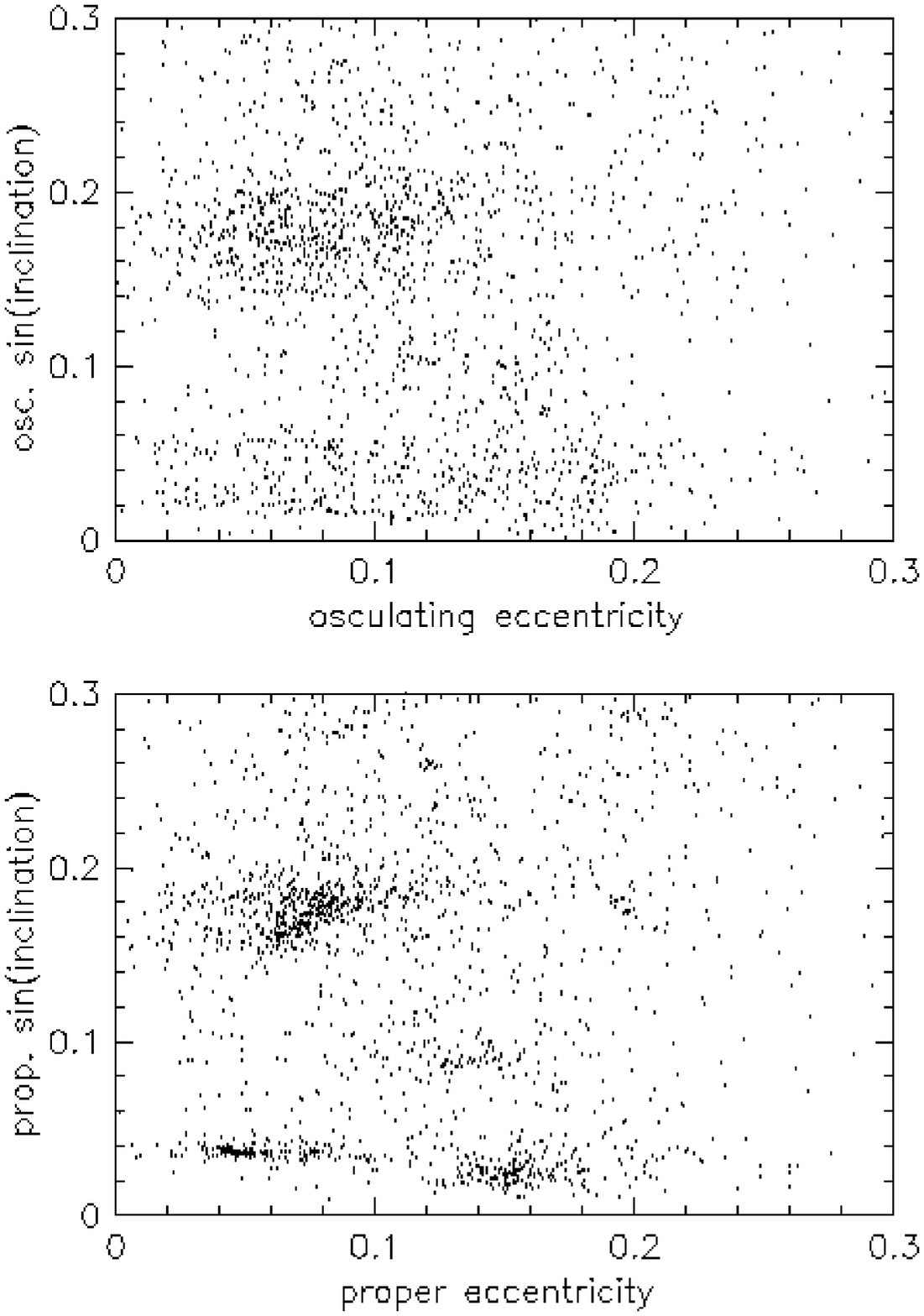}{12cm}{0}{70}{70}{-220}{-50}
\caption{
The dots show the distribution of 1,720 asteroids from the outer region of the main
asteroid belt (proper semi-major axis larger than 2.84 AU). The top panel is constructed with
osculating elements, and the bottom panel with proper elements. The clustering is much stronger
in proper element space.
\label{fig1}
}
\end{figure}

\begin{figure}
\plotfiddle{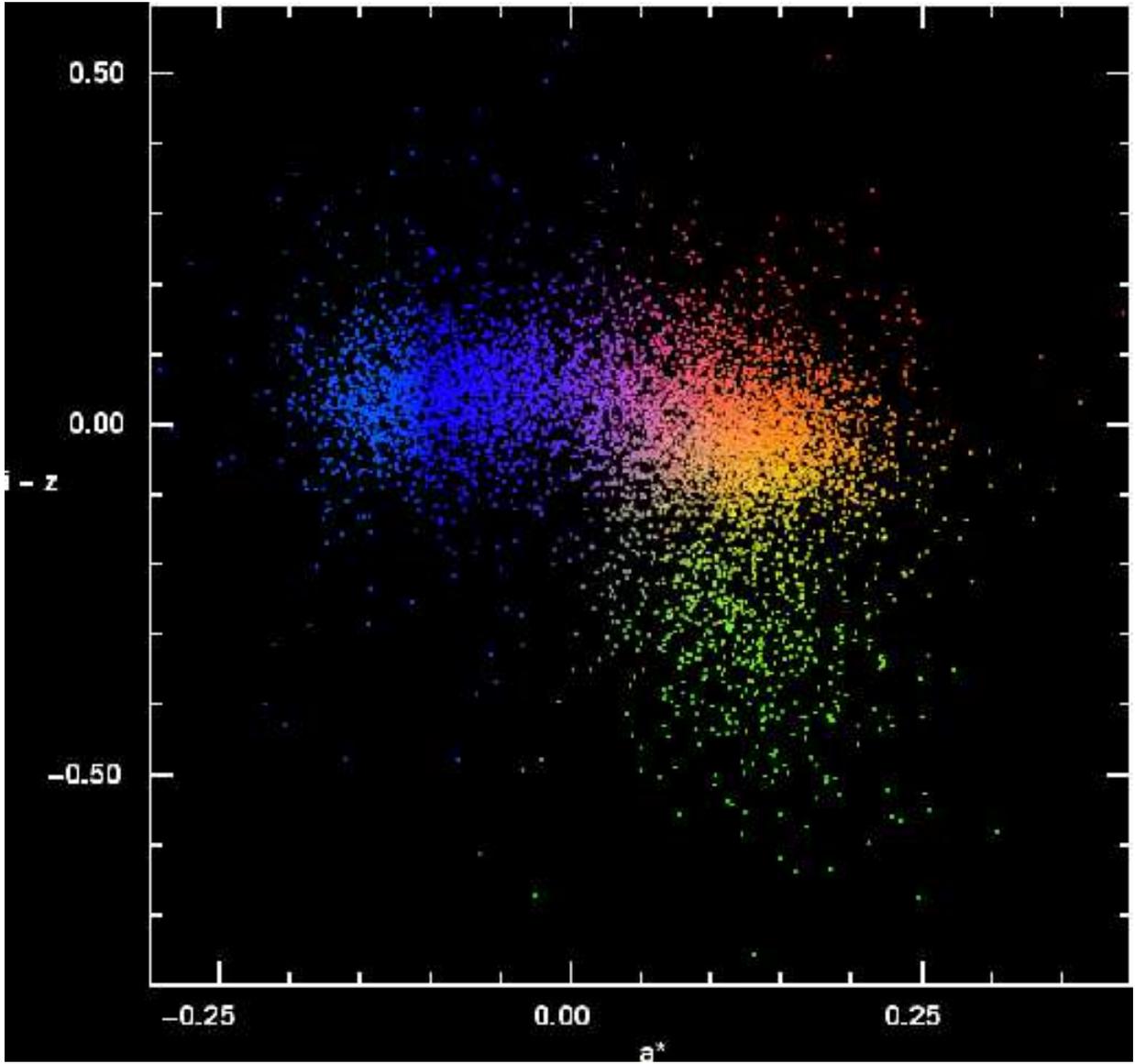}{11cm}{0}{90}{90}{-270}{-140}
\caption{
The dots show the distribution of 6,612 asteroids with available proper orbital
elements in the space spanned by SDSS colors.
\label{fig2}
}
\end{figure}

\begin{figure}
\plotfiddle{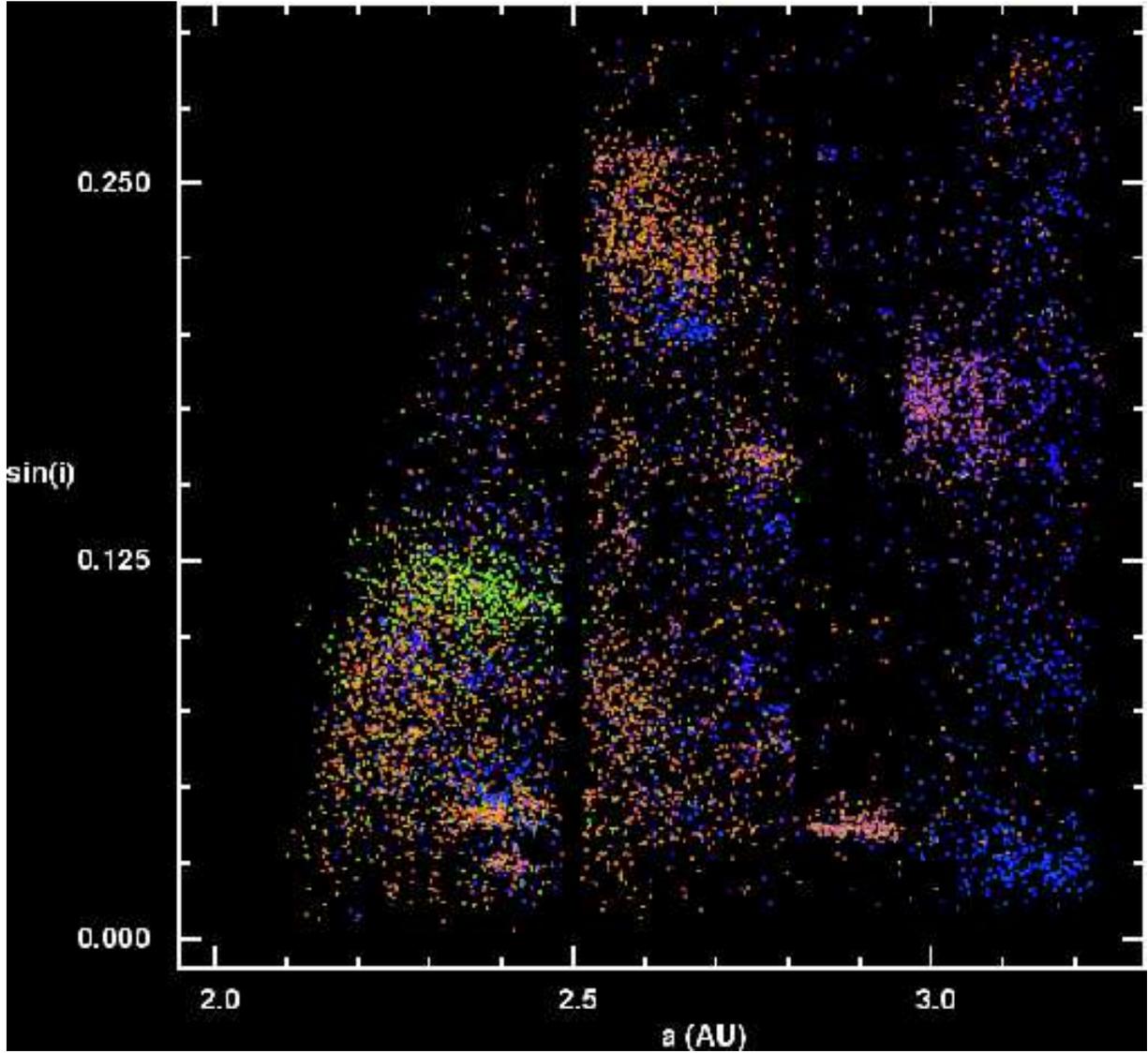}{11cm}{0}{90}{90}{-270}{-140}
\caption{
The dots show the distribution of 6,612 asteroids with available proper orbital
elements in the space spanned by the proper inclination and semi-major axis.
The dots are colored according to their position in the SDSS color-color diagram shown
in Figure~2.
\label{fig3}
}
\end{figure}

\begin{figure}
\plotfiddle{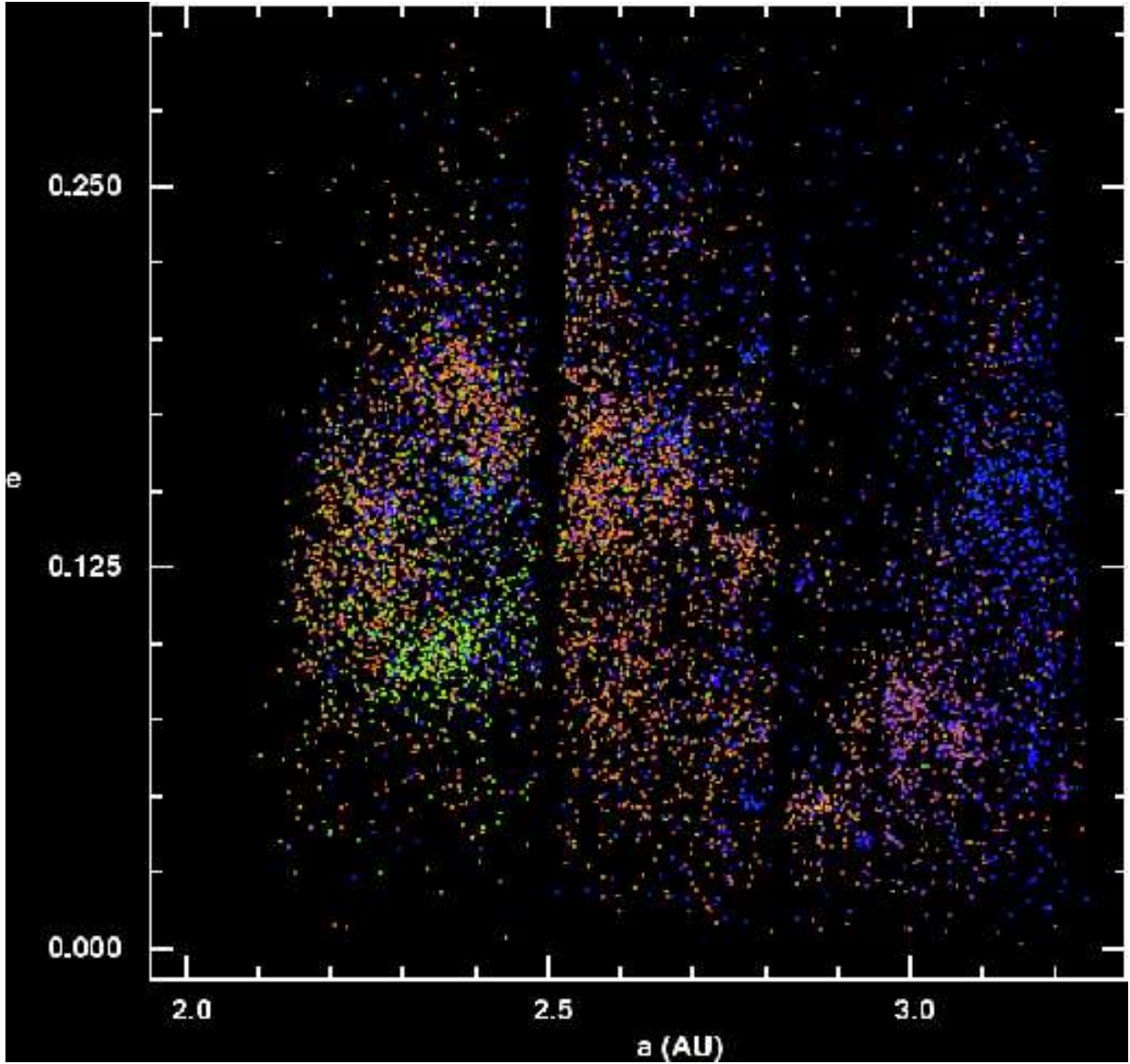}{11cm}{0}{90}{90}{-270}{-140}

\caption{
Same as Figure~3, except that here the distribution in proper eccentricity vs.
semi-major axis is shown.
\label{fig4}
}
\end{figure}

\begin{figure}
\plotfiddle{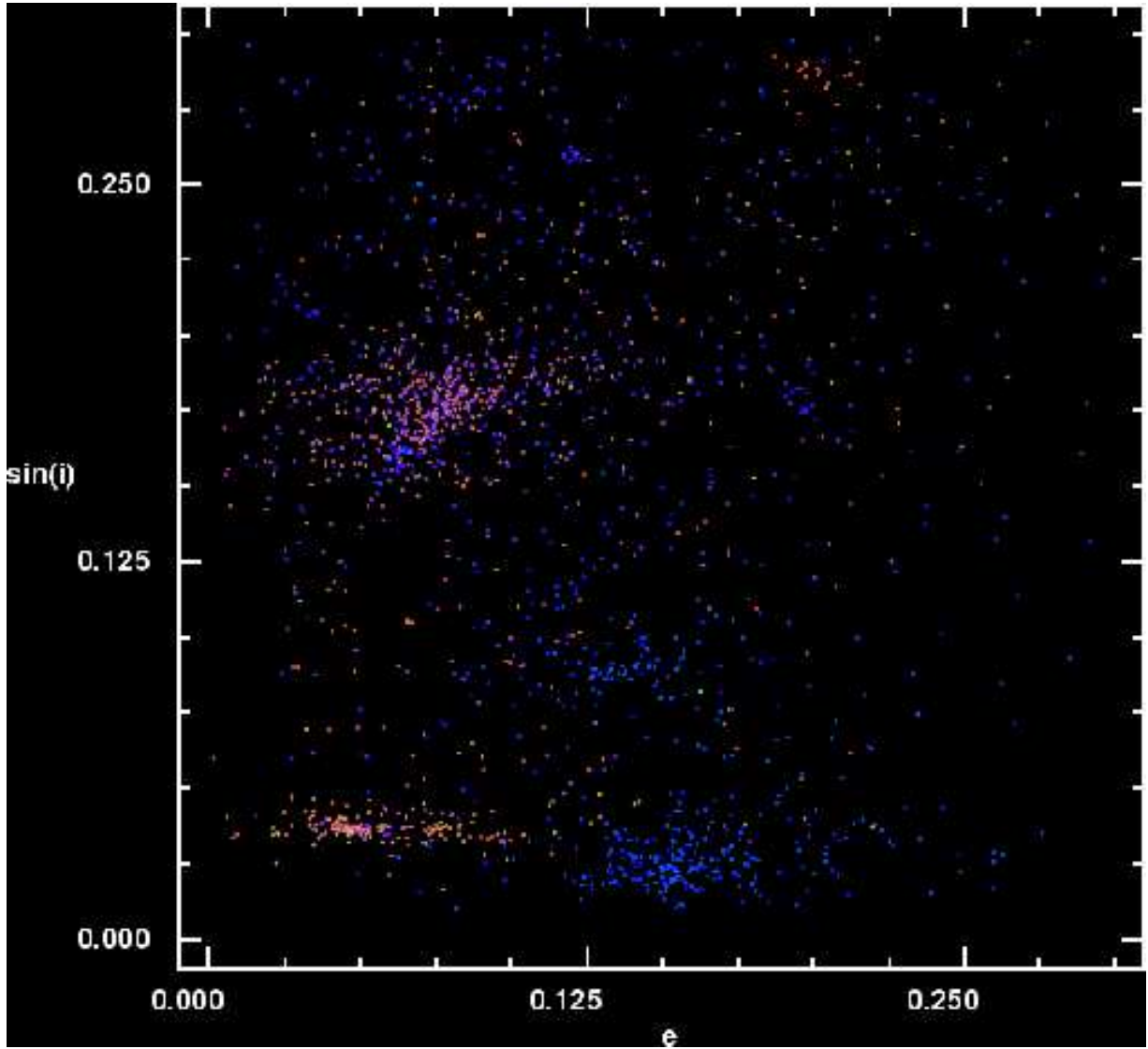}{11cm}{0}{90}{90}{-270}{-140}
\caption{
This is an analogous diagram to that shown in the bottom panel in Figure~1, except
that here the SDSS color information is also displayed, using the color-coding shown
in Figure~2 (only objects with proper semi-major axis larger than 2.84 AU are displayed).
\label{fig5}
}
\end{figure}

\begin{figure}
\plotfiddle{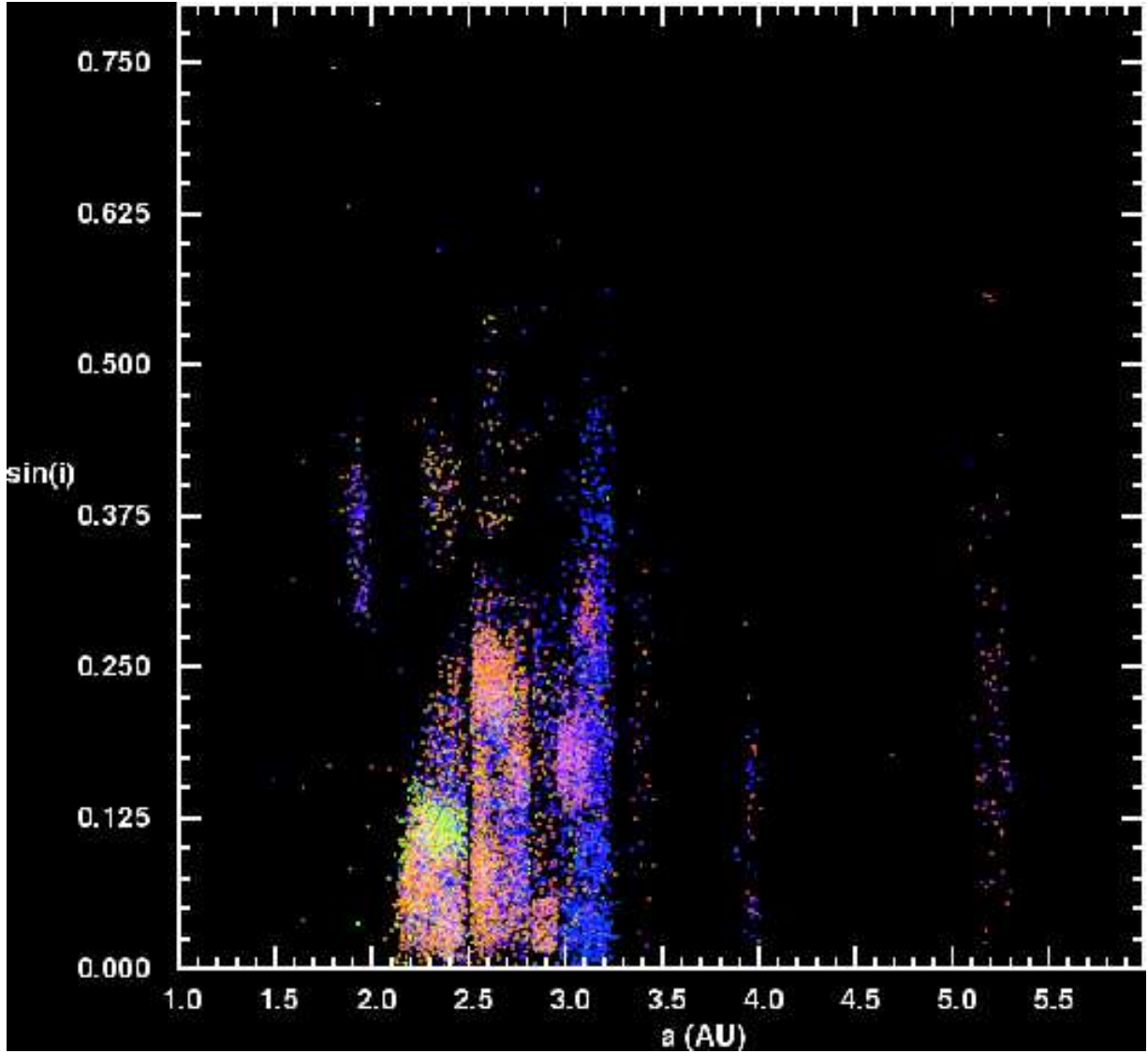}{11cm}{0}{90}{90}{-270}{-140}
\caption{
The dots show the distribution of 10,592 known asteroids observed by the SDSS
in the space spanned by the osculating inclination and semi-major axis.
The dots are colored according to their position in SDSS color-color diagram shown
in Figure~2. Note that the asteroid population is dominated by families.
\label{fig6}
}
\end{figure}

\end{document}